\documentclass[twoside,twocolumn,9pt]{article}
\usepackage{extsizes}
\usepackage[super,sort&compress,comma]{natbib} 
\usepackage[version=3]{mhchem}
\usepackage[left=1.5cm, right=1.5cm, top=1.785cm, bottom=2.0cm]{geometry}
\usepackage{balance}
\usepackage{mathptmx}
\usepackage{gensymb}
\usepackage{sectsty}
\usepackage{graphicx} 
\usepackage{lastpage}
\usepackage[format=plain,justification=justified,singlelinecheck=false,font={stretch=1.125,small,sf},labelfont=bf,labelsep=space]{caption}
\usepackage{float}
\usepackage{fancyhdr}
\usepackage{fnpos}
\usepackage[english]{babel}
\addto{\captionsenglish}{
  
}
\usepackage{array}
\usepackage{droidsans}
\usepackage{charter}
\usepackage[T1]{fontenc}
\usepackage[usenames,dvipsnames]{xcolor}
\usepackage{setspace}
\usepackage[compact]{titlesec}
\usepackage{hyperref}

\usepackage{epstopdf}

\definecolor{cream}{RGB}{222,217,201}

\newcommand{\TiC}{{Ti$_3$C$_2$T$_2$}}
\newcommand{\VC}{{V$_2$CT$_2$}}
\newcommand{\TiCO}{{Ti$_3$C$_2$O$_2$}}
\newcommand{\TiCF}{{Ti$_3$C$_2$F$_2$}}
\newcommand{\VCO}{{V$_2$CO$_2$}}
\newcommand{\VCF}{{V$_2$CF$_2$}}
\begin{document}

\pagestyle{fancy}

\fancyhf{}
\fancyfoot[C]{\thepage}
\renewcommand{\headrulewidth}{0pt}
\renewcommand{\footrulewidth}{0pt}

\makeFNbottom
\makeatletter
\renewcommand\LARGE{\@setfontsize\LARGE{15pt}{17}}
\renewcommand\Large{\@setfontsize\Large{12pt}{14}}
\renewcommand\large{\@setfontsize\large{10pt}{12}}
\renewcommand\footnotesize{\@setfontsize\footnotesize{7pt}{10}}
\makeatother

\renewcommand{\thefootnote}{\fnsymbol{footnote}}
\renewcommand\footnoterule{\vspace*{1pt}%
\color{cream}\hrule width 3.5in height 0.4pt \color{black}\vspace*{5pt}} 
\setcounter{secnumdepth}{5}

\makeatletter 
\renewcommand\@biblabel[1]{#1}            
\renewcommand\@makefntext[1]%
{\noindent\makebox[0pt][r]{\@thefnmark\,}#1}
\makeatother 
\renewcommand{\figurename}{\small{Fig.}~}
\sectionfont{\sffamily\Large}
\subsectionfont{\normalsize}
\subsubsectionfont{\bf}
\setstretch{1.125} 
\setlength{\skip\footins}{0.8cm}
\setlength{\footnotesep}{0.25cm}
\setlength{\jot}{10pt}
\titlespacing*{\section}{0pt}{4pt}{4pt}
\titlespacing*{\subsection}{0pt}{15pt}{1pt}

\fancyhead{}
\renewcommand{\headrulewidth}{0pt} 
\renewcommand{\footrulewidth}{0pt}
\setlength{\arrayrulewidth}{1pt}
\setlength{\columnsep}{6.5mm}
\setlength\bibsep{1pt}

\makeatletter 
\newlength{\figrulesep} 
\setlength{\figrulesep}{0.5\textfloatsep} 

\newcommand{\topfigrule}{\vspace*{-1pt}%
\noindent{\color{cream}\rule[-\figrulesep]{\columnwidth}{1.5pt}} }

\newcommand{\botfigrule}{\vspace*{-2pt}%
\noindent{\color{cream}\rule[\figrulesep]{\columnwidth}{1.5pt}} }

\newcommand{\dblfigrule}{\vspace*{-1pt}%
\noindent{\color{cream}\rule[-\figrulesep]{\textwidth}{1.5pt}} }

\makeatother




\twocolumn[
\begin{@twocolumnfalse}
\begin{center}

{\LARGE \textbf{Influence of stacking, coordination, and surface chemistry on Al intercalation in \VC\ and \TiC\ MXenes for Al-ion batteries}}

\vspace{1em}

Amal Raj Veluthedath Nair$^{a}$,  Nuala M. Caffrey$^{a}$

\end{center}

\vspace{1.5em}

\begin{abstract}

As the energy storage ecosystem evolves beyond lithium, MXenes, a versatile family of 2D materials derived from MAX phases, have emerged as promising candidates for next-generation energy storage electrodes due to their tunable surface chemistry, large interlayer spacing, and excellent electronic conductivity. In this work, we use density functional theory to investigate Ti$_3$C$_2$ and V$_2$C MXenes as cathodes in Al-ion batteries. Four stacking configurations of the two-dimensional sheets and two different ion coordination sites are evaluated to understand their influence on ion intercalation and mobility. We find that the stacking configuration and surface chemistry critically impact interlayer spacing and electrochemical performance. O-terminated layers in an octahedral stacking exhibit remarkable structural stability with minimal interlayer expansion upon ion intercalation, particularly for Al intercalation in V$_2$C which exhibits an interlayer expansion of 0.1~\AA{}, consistent with experimental findings. While octahedral stacking is observed to be energetically more favourable, it reduces ion mobility compared to prismatic stacking. Furthermore, O-terminated MXenes exhibit high theoretical specific capacities, reaching more than 270~mAh/g. F-terminated MXenes are considerably more unstable after intercalation and as a result exhibit much lower Al capacities. These findings highlight the importance of stacking configurations, termination and intercalant chemistry in MXenes for battery applications.
\end{abstract}
\vspace{2em}

\end{@twocolumnfalse}
]

\renewcommand*\rmdefault{bch}\normalfont\upshape
\rmfamily
\section*{}
\vspace{-1cm}

\footnotetext{\textit{$^{a}$~School~of~Physics,~University~College~Dublin,~Dublin~4,~Ireland;~E-mail:~nuala.caffrey@ucd.ie}}
\footnotetext{\textit{$^{b}$~Centre for Quantum Engineering, Science, and Technology, University College Dublin, Dublin 4, Ireland.}}

\section{Introduction}
Lithium-ion batteries (LIBs) are the dominant technology in the battery market, outperforming other battery types in terms of energy density, power density, cycle life, and self-discharge, while also avoiding memory effects.

However, LIBs also suffer from certain limitations, including the high cost per unit energy stored, limited cycle life, voltage degradation, oxidation, and safety risks associated with thermal runaway.~Additional challenges arise from the uneven global distribution of scarce lithium resources, as well as the significant environmental and human costs of lithium mining~\cite{chen2021review}.

Global technological trends are driving the need for energy systems with even greater performance capabilities. 
The European Union's Battery Strategy Action Plan and the Battery 2030+ Roadmap set ambitious targets for energy storage technologies, with a steady scale-up towards the TWh range~\cite{edstrom2020battery} and a target cell-level energy density of at least 500 Wh/kg~\cite{niu2024strategies}. Meeting these targets will require fundamental breakthroughs in battery systems, with improvements in specific energy, volumetric energy density, cycle life, charge/discharge rates, stability, and safety all considered essential.

Alternative charge carries such as monovalent Na and K, as well as multivalent Mg, Ca, Zn, and Al, have the potential to alleviate pressure on lithium resources and, in some cases, surpass the performance of LIBs.
~Like lithium, these systems typically rely on the reversible intercalation of metal ions into host electrode materials, a process that enables charge storage and release.
Among these, sodium-ion batteries (SIBs) are considered particularly promising as either a complement to, or replacement for, LIBs, due to their compatibility with similar electrode materials and electrolyte formulations, which would facilitate a relatively straightforward  transition~\cite{chayambuka2020li}.~In addition, sodium is more abundant and presents a lower risk of thermal runaway. However, its larger atomic radius and higher specific weight result in a lower theoretical capacity and energy density, limiting the applicability of SIBs to certain use cases~\cite{durmus2020side}.

The use of Mg and Al as pure metal anodes offers the potential for both high gravimetric and volumetric capacities. For Mg, the theoretical gravimetric and volumetric capacities are 2205 mAh~g$^{-1}$ and 3833 mAh~cm$^{-3}$ respectively -- substantially higher than those of the graphite anodes used in LIBs, which only deliver 372 mAh~g$^{-1}$ and 800 mAh~cm$^{-3}$~\cite{canepa2017odyssey}.~Furthermore, magnesium is at least four times abundant in the Earth's crust than lithium.~It also offers advantages such as safer handling, a reduced propensity for dendrite formation, and the avoidance of toxic by-products~\cite{yoo2013mg}. 

Research into Al-ion batteries (AIBs) remains at an early stage; however, the use of an aluminium metal anode offers even higher theoretical gravimetric and volumetric capacities of 2980 mAh~g$^{-1}$ and 8040 mAh~cm$^{-3}$, making it one of the most promising candidates among all metal-ion battery systems~\cite{elia2016overview}.~In addition, the abundance of aluminium in the Earth's crust would contribute to significantly lower production costs compared to lithium-based systems~\cite{xu2021engineering}. 

Although the reversible intercalation of Mg, Ca, and Al metal ions could theoretically enable capacities double or even triple those of LIBs, practical challenges related to electrolyte and cathode selection currently limit the maximum achievable energy density~\cite{el2020exploits}.
The main drawbacks of existing electrode materials include significant volume expansion upon ion intercalation, which leads to structural degradation; strong host-guest interactions that hinder ion mobility; intrinsic hydrophobicity; and low electrical conductivity~\cite{rong2015materials}.

Layered MXene materials are promising candidates as electrodes for post-LIB technologies. MXenes are a family of layered two-dimensional (2D) transition metal carbides, nitrides and carbonitrides that feature large interlayer spacings and high surface areas. They also exhibit electrical conductivities comparable to graphene, excellent thermal stability, stiffness surpassing that of steel, and rich surface chemistries~\cite{naguib2023two}.

Experimental studies have demonstrated that MXenes can deliver high volumetric capacities when intercalated with a variety of monovalent and multivalent cations ~\cite{lukatskaya2013cation,tang20182d,kajiyama2016sodium}.
Among them, V\textsubscript{2}C stands out as showing particular promise as an Al-ion battery cathode, having demonstrated a high reversible capacity attributed to Al\textsuperscript{3+} ion intercalation between its layers~\cite{vahidmohammadi2017two} due to the remarkably minor increase in interlayer spacing during the intercalation process.
Furthermore, exceptional specific capacities exceeding 300 mAh~g$^{-1}$ at high charge/discharge rates were demonstrated in delaminated \VC. Despite these impressive electrochemical performances -- among the best reported for aluminium-ion battery cathodes -- important questions remain unresolved at the atomic scale. In particular, the detailed mechanism of Al$^{3+}$ intercalation, the origin of a gradual capacity fade observed during cycling, and the factors controlling ion transport and structural stability remain to be fully understood. 

In addition to their compositional diversity, MXenes exhibit variations in their interlayer stacking arrangements and surface terminations, both of which can strongly influence ion intercalation and charge storage behavior. 
Surface terminations --- typically --O, --OH, or --F groups introduced during synthesis --- modify the local electronic structure and binding characteristics of the MXene surfaces~\cite{saha2021enhancing,kajiyama2017enhanced,xie2014role}, while the stacking of adjacent layers determines the accessible interlayer spacing and diffusion pathways for intercalating ions~\cite{wang2016resolving}. 

In this study, we investigate how these structural and surface termination factors affect ion intercalation in two representative multilayered MXenes: Ti\textsubscript{3}C\textsubscript{2}, the prototypical and most extensively studied MXene, and V\textsubscript{2}C, which, as previously mentioned, has shown promise as a Al-ion battery cathode~\cite{vahidmohammadi2017two}. Aluminium-ion intercalation is of primary interest, but for comparison and broader insight, we also examine Na and Mg intercalation, to explore how intercalation energetics evolve from monovalent Na\textsuperscript{+} to trivalent Al\textsuperscript{3+}. Using density functional theory, we access atomic-scale properties and mechanisms that are challenging to resolve experimentally, allowing us to clarify how stacking configurations, surface and ion chemistry collectively influence electrochemical performance.

\section{Crystal Structure}

The crystal structure of MXenes is inherited from their parent MAX phase compounds, which crystallize in the hexagonal space group P6$_3$/mmc~\cite{sokol2019chemical}.
MXenes are produced by the selective chemical etching of the relatively weakly bound A layer from a MAX phase material, where M is an early transition metal atom, A is a group A element, and X is carbon and/or nitrogen.~All ternary MAX phases have the general formula M$_{n+1}$AX$_n$ ($n$ = 1$-$4).

\begin{figure}[ht!]
    \centering
    \includegraphics[width=\columnwidth]{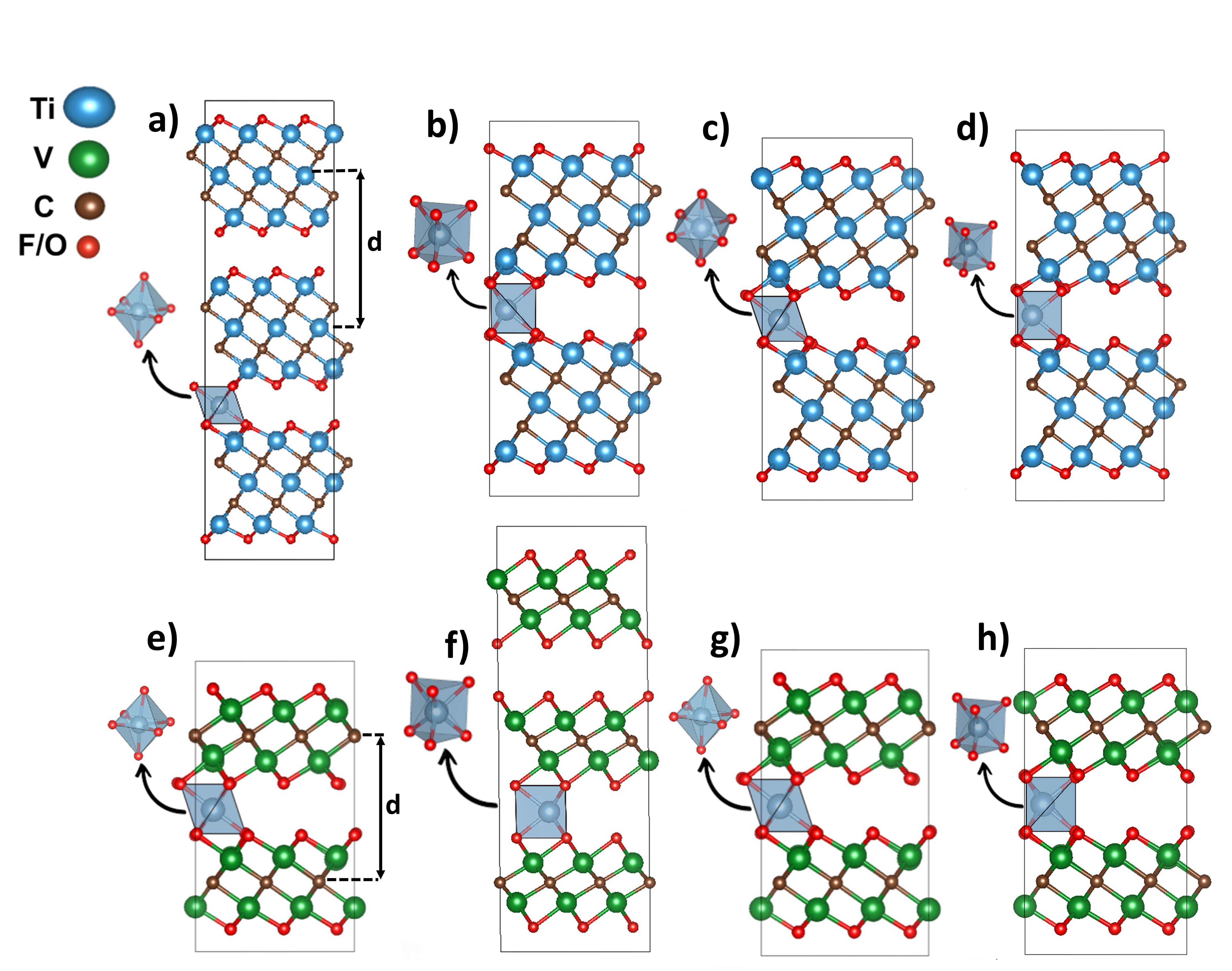}  
    \caption{(a) WS--oct \TiC, (b) WS--pris \TiC, (c) ZZ--oct \TiC, and (d) WS--pris \TiC. (e), (f), (g), and (h) present the corresponding \VC\ structures.}
    \label{fig1}
\end{figure}

The etching process typically employs a variety of etchants, including HF, HCl, HBr, and HI, resulting in MXene surfaces terminated with functional groups such as F, O, OH, Cl, and Br. Consequently, the chemical formula of MXenes is often represented as M$_{n+1}$X$_n$T$_x$, where T$_x$ denotes a mixture of these surface terminations~\cite{natu2023mxene}. The nature and composition of these terminations depends on a number of factors including the M and X elements, the number of layers ($n$), the etching conditions, and the type and concentration of the etchant~\cite{anayee2020role}. As terminations in experiment are mixed and random, modelling them becomes complex. To avoid this complexity, our study is restricted to uniform  terminations of oxygen and flourine.

If there is no relative movement of adjacent layers during chemical etching, the resulting stacking is trigonal prismatic (space group P6$_3$/mmc), in which the surface terminations of successive layers align directly on top of each other, forming triangular prisms, as illustrated in Figs.~\ref{fig1}(b) and (d). This arrangement, experimentally observed for certain MXenes and termination types~\cite{wang2015atomic, kamysbayev2020covalent}, will be referred to as the `pris' stacking in this work.  

Alternatively, octahedral stacking arises when every second layer of the pris stacking shifts by one-third of the unit cell length along the [$110$]-direction, causing the terminating groups of adjacent layers to form octahedral coordination environments. This stacking also belongs to the P6$_3$/mmc space group and will be termed `oct' stacking here. It is shown in Figs.~\ref{fig1}(a) and (c).
Experimental evidence suggests that oct stacking occurs more frequently than pris stacking~\cite{li2019element,zhang2019atomic}.

Moreover, several studies report the coexistence of mixed pris and oct stacking within the same MXene sample, indicating that subtle variations in the etching process may determine the resulting stacking configuration~\cite{wang2015atomic,  halim2014transparent}.

In addition to the stacking types formed by sliding one layer relative to the next, an alternative stacking configuration arises when alternate MXene layers are rotated by 180\degree\ around the $c$-axis. This arrangement has been termed whip-stitch (WS) stacking by Hadler-Jacobsen et al.~\cite{hadler2022importance}, to distinguish it from the more common zig-zag (ZZ) stacking observed in MXenes and their parent MAX phases. Within the WS stacking, both prismatic and octahedral arrangements of the terminating groups are possible, with transitions between pris and oct stackings achievable through layer gliding. Some limited experimental evidence supports the existance of WS stacking~\cite{halim2014transparent}, including observations of unclear or mixed stacking patterns~\cite{lu2019ti}.
In this work, these configurations are designated as WS--pris and WS--oct,  distinct from the ZZ--pris and ZZ--oct configurations (compare Fig.~\ref{fig1} (a) and (c)). WS--pris corresponds to AA-type stacking, ZZ--pris and ZZ--oct correspond to AB-type stacking, and WS--oct corresponds to ABC-type stacking -- terminology that has occasionally been used in the literature to differentiate these stacking variants. 

Intercalation can alter the relative stability of MXene stacking configurations, with these changes depending on the nature of the intercalant.~For instance, Wang~\textit{et al.} experimentally observed a transition from ZZ--pris to ZZ--oct stacking upon Al intercalation using annular bright-field (ABF) imaging~\cite{wang2015atomic}.~Such stacking changes must be considered when evaluating MXenes as battery electrodes.
\section{Theoretical Methods}

All calculations were performed using Density Functional Theory (DFT) as implemented in the Vienna \textit{ab initio} Simulation Package (VASP, version 6.4.4.)~\cite{kresse1996efficient}. Valence electrons and ion core interactions were treated using the projector-augmented wave (PAW) method~\cite{kresse1999ultrasoft}. The Perdew, Burke, and Ernzerhof (PBE) parameterization of the generalized gradient approximation (GGA) was employed to describe the exchange-correlation functional~\cite{perdew1996generalized}.~van der Waals interactions were accounted for using the D3 dispersion correction scheme of Grimme~\cite{grimme2010consistent}, chosen based on its demonstrated reliability in previous studies on MXene layers~\cite{tygesen2019role,hadler2021stacking,hadler2022importance}.

A plane-wave energy cut-off of 500~eV was used. Electronic self-consistency was achieved with an energy convergence criterion of $1\times$$10^{-6}$~eV, while ionic relaxations were converged until forces were below 0.001~eV/\AA{} for all stackings, except for the Al intercalated ZZ-pris \VCF\, where a less strict criterion of 0.01~eV/\AA{} was applied. The tetrahedron method with Bl\"{o}chl corrections was applied for Brillouin zone intergrations~\cite{blochl1994improved}.

The Brillouin zone sampling employed Monkhorst-pack \textit{k}-point meshes, whose densities varied according to the stacking type and size of the simulation cell. Specifically, a $12\times12$ mesh was used in the $ab$-plane, combined with 2, 4, 4 and 8 $k$-points along the $c$-axis for WS--oct, ZZ--oct, ZZ--pris, and WS--pris stacking configurations, respectively. (Note that the WS--oct stacking contains three MXene layers per unit cell, as shown in Fig~\ref{fig1} (a), while both ZZ-oct and ZZ--pris have two layers per unit cell, and WS--pris can be described with a single layer per unit cell.) 
These simulation parameters yield structural and electronic properties for Ti$_3$AlC$_2$ and V$_2$AlC MAX phases consistent with experimental data~\cite{pietzka1994summary,hamm2018structural} and previous DFT studies~\cite{zhou2001electronic,pinek2018electronic}. 

To model increasing intercalant concentrations in ZZ--oct and ZZ--pris stacked structures, 3$\times$3$\times$1 supercells were used, with a correspondingly reduced \textit{k}-point mesh of $4\times4$ in the $ab$-plane. 
The lowest dilute limit corresponded to a single Al in one of the two interlayer spacings of the supercell, equating to a concentration of 0.055 Al per formula unit. 
Higher concentrations are modeled by sequentially adding one additional intercalant per interlayer spacing. The most energetically favorable site of those available in the chosen supercell was found to be the one farthest from previous intercalants.

Open circuit voltage (OCV) calculations were performed assuming the following reaction between two states of alumination:
$$(x_2 - x_1)\mathrm{Al}^{3+} + 3(x_2 - x_1)e^- + \mathrm{Al}_{x_1}\mathrm{MXene} \rightleftharpoons \mathrm{Al}_{x_2}$$

where $x_2 > x_1$, and bulk aluminum in a face-centered cubic (fcc) lattice serves as the reference.
The average equilibrium voltage~\cite{urban2016computational} at alumination state $x$ is given by the Gibbs free energy change of the reaction as:
\begin{multline}
V(x)  = -\frac{\Delta G}{z F} \approx \\[10pt]
   \hspace{0.5cm} -\frac{E(\mathrm{Al}_{x_2}\mathrm{MXene}) - E(\mathrm{Al}_{x_1}\mathrm{MXene})-(x_2 - x_1)E(\mathrm{Al})}{3(x_2 - x_1)F}
\end{multline}

Here, $z$ is the number of electrons transferred per Al ion ($z = 3$) and $F$ is the Faraday constant. Entropic and volumetric contributions to $\Delta G$ were neglected, as they are typically small at room temperature~\cite{aydinol1997ab}. Thus, $\Delta G$ is approximated by the formation energy, defined as:
$$
E_{\mathrm{formation}} = E(\mathrm{Al}_x\mathrm{MXene}) - E(\mathrm{MXene}) - xE(\mathrm{Al})
$$
with the average formation energy per Al given by: $\overline{E}_{\mathrm{formation}}$,

$$
   \overline{E}_{\mathrm{formation}} =  \frac{E(\mathrm{Al}_x\mathrm{MXene}) - E(\mathrm{MXene}) - xE(\mathrm{Al})}{x}
$$

Migration barriers for Al ions in the dilute limit (one Al per $3\times3$ unit cell) were calculated using the Nudged Elastic Band (NEB) method~\cite{henkelman2000improved} for the ZZ--oct and ZZ--pris stacking configurations of \TiCO\ and \VCO. Forces were relaxed to below 0.01 eV/\AA{} using the Quick-Min optimizer. Initial and final states of migration correspond to the most stable site and a neighbouring metastable site, respectively. For trigonal prismatic stacking, the metastable site has trigonal prismatic coordination with the terminating groups and is located between two Ti atoms. For octahedral stacking, the site lies inside a tetrahedral void formed between a termination group and Ti atoms. Two intermediate images were used in the NEB calculations to generate the migration energy barrier profile, which was then symmetrized between adjacent carbon atoms.

Atomic structures were plotted using the VESTA program~\cite{momma2011vesta}.

\section{Results and discussion}
\subsection{Unintercalated MXenes}
Several computational studies have traditionally focused solely on the ZZ–pris stacking configuration without exploring alternative stackings. Although this approach simplifies analysis, it may overlook important structural rearrangements induced by intercalation, potentially limiting the accuracy of predicted electrochemical properties. Here we have analysed the relative structural stability of four stacking configurations.
Table~\ref{tab:Tab1} summarizes the relative energy differences between the four stacking configurations considered for O and F-terminated \TiC\ and \VC. It is calculated as:
$\Delta E_{\mathrm{stacking}}$ = $E_{\mathrm{stacking}}$ - $E_{\mathrm{most~stable~stacking}}$. 
In this table, a value of zero corresponds to the most stable stacking configuration for each MXene, and positive values indicate the energy difference relative to that minimum.

\begin{table}[h]
\small
\caption{Energy differences ($\Delta E_{\mathrm{stacking}}$) between different stacking configurations of --O and F-terminated \TiC\ and \VC. Units are meV per formula unit (meV/f.u.). The lowest energy stacking configuration is set to zero for each MXene and termination, with other values relative to this minimum.}
  \label{tab:Tab1}
  \begin{tabular*}{0.48\textwidth}{@{\extracolsep{\fill}}lcccc}
    \hline
     & ZZ--pris & ZZ--oct & WS--pris & WS--oct \\
    \hline
    Ti$_3$C$_2$O$_2$ & 89 & 12 & 90 & \textbf{0} \\
    Ti$_3$C$_2$F$_2$ & 80 & 8  & 79 & \textbf{0} \\
    V$_2$CO$_2$      & 52 & 2  & 50 & \textbf{0} \\
    V$_2$CF$_2$      & 66 & 4  & 63 & \textbf{0} \\
    \hline
  \end{tabular*}
\end{table}

Our results show that the WS--oct stacking configuration is consistently the lowest in energy for all MXenes studied. The energy difference between WS--oct and the ZZ--oct stacking is small, not exceeding 12~meV/f.u., and is as low as 2~meV/f.u.\ for \VCO. 
These differences are significantly below thermal energy at room temperature, indicating that both stackings may coexist thermally. 
Similarly, the energy differences between ZZ--pris and WS--pris stackings are very small, less than 3~meV/f.u.\ in all cases, suggesting a near degeneracy.

As transitioning from ZZ to WS stacking requires a 180$^\circ$ rotation combined with layer sliding, kinetic barriers are likely responsible for the predominance of ZZ stacking observed experimentally~\cite{halim2014transparent,wang2015atomic,cheng2018understanding,kamysbayev2020covalent}, despite the slightly higher energy compared to WS. 

However, the energy difference between the oct and pris stacking is substantially larger in all cases. For \TiC, this difference ranges between 72 and 90~meV/f.u., depending on termination and stacking type, while for \VC\ it is somewhat smaller, ranging from 50 to 63~meV/f.u. This clearly indicates that octahedral stacking is energetically preferred over prismatic stacking in unintercalated MXenes.

\subsection{Intercalated MXenes - Dilute Limit}
Due to an experimental prevalence, we now focus on the ZZ-type stacking type. For both ZZ-oct and ZZ-pris stacking, Na, Mg and Al were intercalated at a dilute concentration of 0.055/f.u.\ with uniform --O and --F terminations.

\begin{figure}[h!]
    \centering
    \includegraphics[width=1\columnwidth]{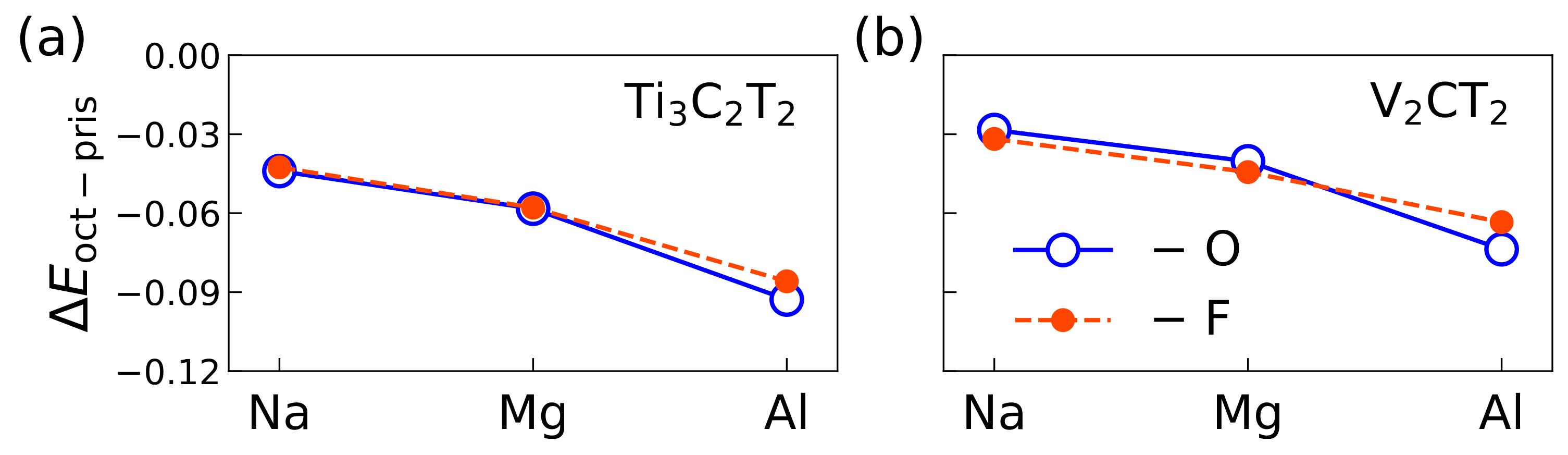}
    \caption{(a) Energy difference between ZZ--oct and ZZ--pris stacking ($\Delta E_\mathrm{oct-pris}$) for \TiC\ (b) Energy difference between ZZ--oct and ZZ--pris stacking for \VC, where T = --O and T = --F terminations are denoted as solid blue and dashed orange lines. Units are meV per formula unit (meV/f.u.)}
    \label{Fig2}
\end{figure}

Fig.~\ref{Fig2} shows the energy difference $\Delta E_{\text{oct--pris}}
 = E_{\mathrm{ZZ-oct}} - E_{\mathrm{ZZ-pris}}$ for (a) \TiC\ and (b) \VC, indicating that ZZ-oct stacking is consistently more stable than ZZ-pris for all intercalants and terminations. The relative stability of ZZ-oct stacking increases as the intercalant is changed from Na to Mg to Al, with Al intercalation producing the largest energy difference (up to 93 meV/f.u.\ for \TiC). The differences are generally higher in \TiC\ than in \VC.

Notably, the energy difference is smallest for Na intercalants -- 44 and 28meV/f.u.\ for \TiCO\ and \VCO, respectively -- which is consistent with the observation that the preferred stacking for Na intercalation remains a point of contention in the literature. Several previous computational studies have reported ZZ-pris as the more stable stacking for Na~\cite{hadler2022importance, caffrey2018effect}, highlighting a strong sensitivity to the choice of exchange-correlation functional and van der Waals corrections. This computational uncertainty mirrors experimental ambiguities regarding the dominant stacking sequence, and likely arises from the small energy differences involved between competing configurations.

The choice of termination (--O vs. --F) has a minor effect on $\Delta E$ except for Al intercalation in \VC, where it increases the relative stability by about 10 meV/f.u.

Formation energies, defined relative to bulk Al and pristine MXenes, can provide further insight into the thermodynamic favorability of these intercalated structures.
For \TiCO\ Al intercalation is thermodynamically favorable in the ZZ--oct stacking in the dilute Al concentration limit, with formation energies of -292~meV/f.u. It is also weakly favourable in the ZZ--pris stacking at -2~meV.
In contrast, \TiCF\ exhibits positive formation energies for both ZZ--pris and ZZ--oct stackings, indicating that even dilute Al intercalation is thermodynamically unfavorable in F-terminated structures.

For \VC, Al intercalation in the ZZ--oct stacking with O-terminations yields a negative formation energy at the lowest Al concentration (-292 meV/f.u.\ at 0.055 Al/f.u.), indicating thermodynamic favorability. In contrast, the corresponding ZZ--pris stacking with O-terminations at the same intercalant concentration is markedly less favorable, with a formation energy of -2 meV/f.u.

For V$_2$CF$_2$, small formation energies (less than -18 meV/f.u.) are obtained across all stackings, indicating substantially reduced thermodynamic favourability. Nevertheless, Al intercalation in \VCF\ remains weakly favourable,  in contrast to \TiCF, for which Al intercalation is always thermodynamically unfavourable.

These results demonstrate that although ZZ--oct stacking is generally energetically preferred upon Al intercalation, the thermodynamic stability of the intercalated structures is governed primarily by the surface termination and the MXene chemistry. In particular, O-terminations promote thermodynamically favourable Al intercalation, whereas F-terminations substantially suppress stability, with a stronger effect in Ti based MXenes than in their V based counterparts. The evolution of formation energy with increasing Al concentration is discussed later.

Structural changes during charge/discharge cycles are a major contributor to electrode degradation. To access these structural changes, we next examine volume variations after intercalation in \TiC\ and \VC, again considering both the ZZ--oct and ZZ--pris stackings.
Intercalation induces only marginal changes in the in-plane lattice parameter, $a$, regardless of the stacking or MXene type. The nature of the termination groups has a minor effect: intercalation into O-terminated MXenes slightly decreases $a$, while F--terminations cause a small increase. In all cases, the percentage change in $a$ remains below 0.5\%.

\begin{figure}
    \centering
    \includegraphics[width=1\columnwidth]{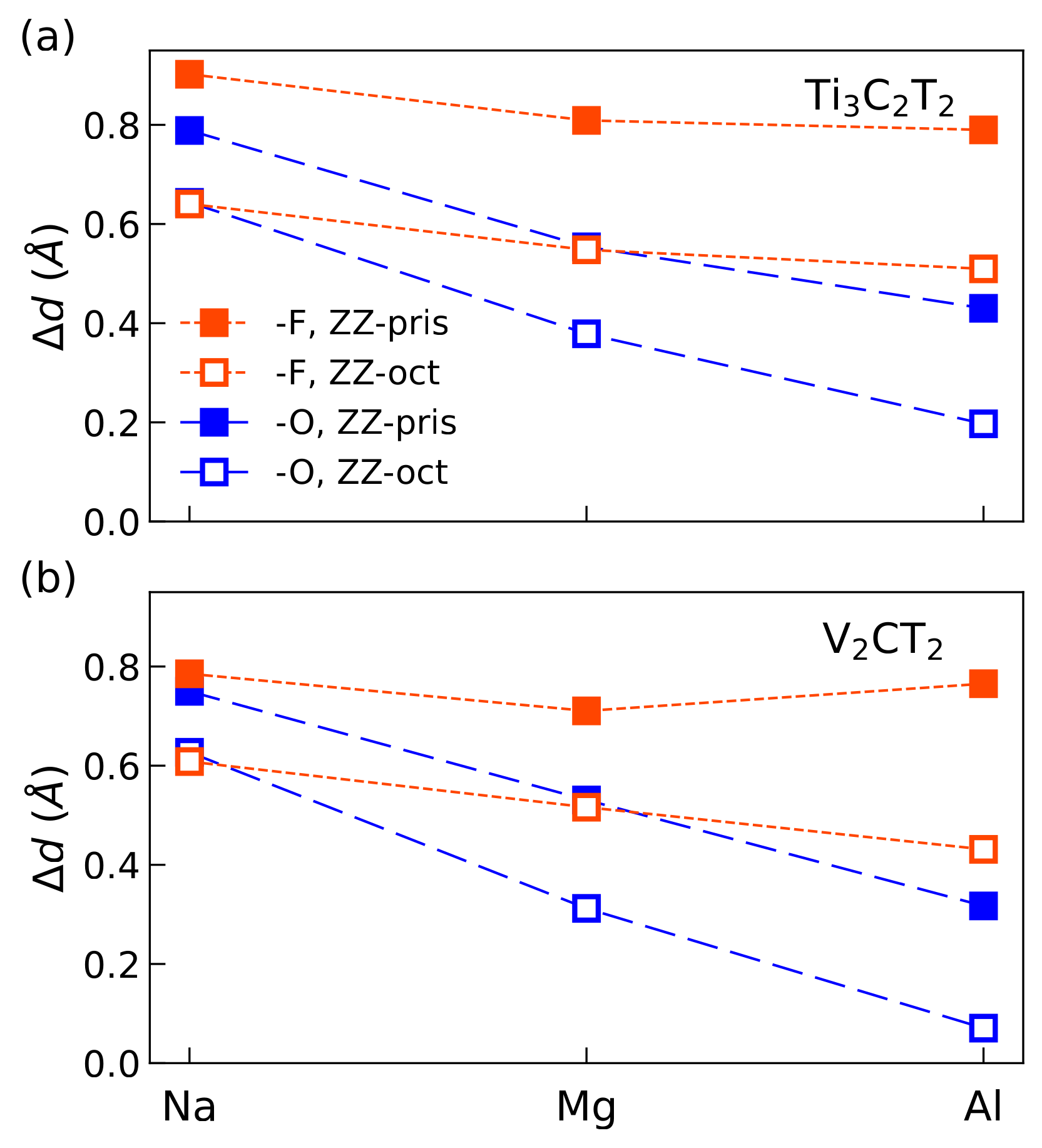}
    \caption{Change in interlayer distances ($\Delta d$) after intercalation of Na, Mg, Al for ZZ--oct and ZZ--pris stacking, denoted as open squares and solid squares, for (a) \TiC\ and (b) \VC. O--terminations as long dashed blue lines and F--terminations as short dashed orange lines.}
    \label{Fig3}
\end{figure}

In contrast, the out-of-plane lattice constant exhibits more substantial changes upon intercalation.~Fig.~\ref{Fig3} shows the change in interlayer distance, $\Delta d$, after intercalation of Na, Mg, or Al into a  3$\times$3 unit cell of (a) \TiC\ and (b) \VC, corresponding to intercalation levels of 0.11/f.u.\ for each ion. Here, $d$ is defined as the distance between the central Ti atoms in \TiC\ or the central C atoms in \VC\ across adjacent layers, as illustrated in Fig.~\ref{fig1}. $\Delta d$ then refers to the change in $d$ relative to the unintercalated ZZ--oct structure. 

Considering first the most stable ZZ--oct stacking (open symbols with long dashed lines), Na intercalation results in the largest $\Delta d$: 0.63~\AA\ for \TiCO\ and 0.64~\AA\ for \VCO.
 
Replacing --O terminations with --F (open orange symbols) does not significantly alter $\Delta d$ for Na intercalation in either MXene.
If, as previously reported, Na intercalation results in a ZZ--pris stacking~\cite{wang2015atomic,kajiyama2017enhanced}, (solid square symbols), $\Delta d$ increases to 0.79~\AA\ for \TiCO\ and 0.75~\AA\ for \VCO. For --F terminations, these values increase slightly to 0.90~\AA\ and 0.78~\AA, respectively (solid orange symbols).

For Mg intercalation in the ZZ--oct stacking, $\Delta d$ reduces to 0.38~\AA\ for \TiCO\ and 0.31~\AA ~for \VCO. Switching to F--terminations again increases $\Delta d$ to 0.55~\AA\ and 0.52~\AA, respectively. 
In the ZZ--pris stacking, Mg intercalation results in $\Delta d$ values of 0.53~\AA\ for \TiCO\ and 0.55~\AA\ for \VCO, with the largest interlayer expansions occurring for the F-terminated pris stacking at 0.80~\AA\ and 0.71~\AA, respectively.

The smallest interlayer expansions are found to occur for Al intercalants in the ZZ--oct stacking. In the case of \TiCO, $\Delta d$ is only 0.12~\AA\ and even smaller, at 0.07~\AA, for \VCO. These values increase if the --O terminations are replaced with F terminations to 0.51~\AA\ and 0.43~\AA\ for \TiCF\ and \VCF\ respectively. 

The change in interlayer distance is larger in the ZZ-pris stackings. 
For \TiCO, $\Delta d$ is 0.43~\AA, while for \VCO\ it is 0.32~\AA. With F-- terminations, $\Delta d$ increases to 0.79~\AA\ for \TiCF\ and to 0.76~\AA\ for \VCF.

These results provide an atomic-scale explanation for experimental observations by VahidMohammadi~\textit{et al.}\cite{vahidmohammadi2017two}, who reported a modest 0.10~\AA\ increase in interlayer spacing upon Al intercalation into \VC, although the stacking order was not explicitly determined in that work. This value is consistent with our calculated $\Delta d$ for the energetically preferred O terminated ZZ--oct structure, reinforcing the conclusion that this stacking dominates at dilute Al concentrations. Nevertheless, direct comparison with experiment should be made cautiously, as co-intercalation of water or other species is likely, and the measurements were performed in the fully discharged state.

\subsection{Stacking Dependence of Al-ion Diffusion Barriers}

\begin{figure}[ht]
    \centering
    \includegraphics[width=1\columnwidth]{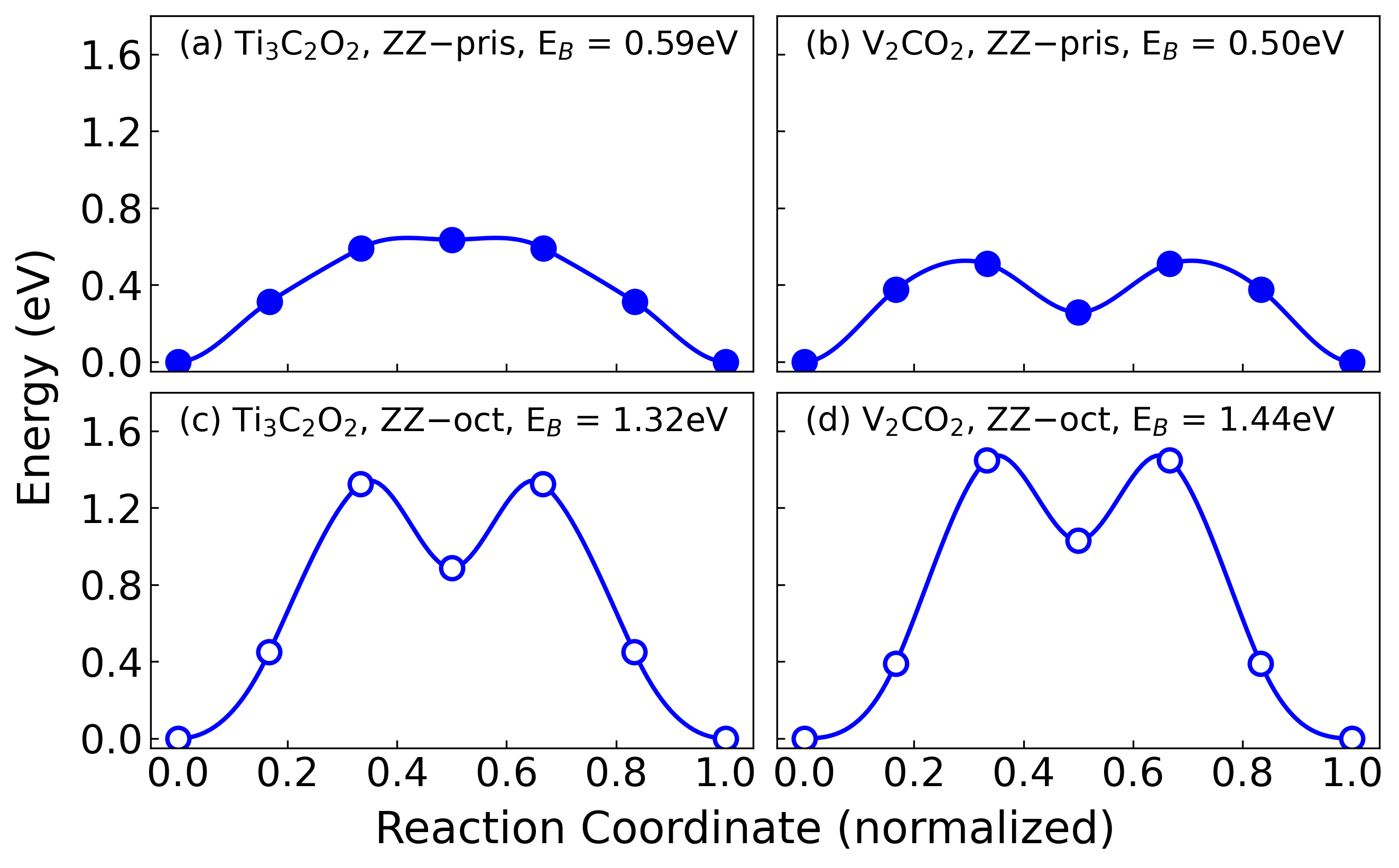}
    \caption{Calculated Al migration energy barriers in (a) ZZ--pris stacked \TiC\ (0.59eV), (b) ZZ--pris stacked \VC\ (0.50eV), (c) ZZ--oct stacked \TiC\ (1.32eV), and (d) ZZ--oct stacked \VC\ (1.44eV).}
    \label{fig:migration-profiles}
\end{figure}

Efficient ion migration is essential for maintaining charge-storage capacity over repeated charge–discharge cycles. In the following, we therefore focus exclusively on O-terminated structures, as F-terminated MXenes were shown in the preceding section to be thermodynamically less favorable for Al intercalation. The calculated migration barriers for Al intercalation in \TiCO\ and \VCO\ are shown in Fig.~\ref{fig:migration-profiles}. 
For both systems, higher migration barriers are obtained in the ZZ--oct stacking than in the ZZ--pris stacking. In the case of \TiCO, the Al migration barrier is 0.59~eV for the ZZ--pris configuration and increases substantially to 1.32~eV in the ZZ--oct configuration. A similar trend is observed for \VCO, where the barrier rises from 0.50~eV in ZZ--pris stacking to 1.44~eV in ZZ--oct stacking. These results indicate that Al migration is energetically more favorable in the ZZ--pris stacking for both materials.

These migration barriers are significantly higher than those reported for other intercalants, including Li, Na, and Mg, in previous studies by Hadler-Jacobsen~\textit{et al.}\cite{hadler2021stacking,hadler2022importance} and for Li and Mg by Kaland~\textit{et al.}\cite{kaland2020mxenes}.

The elevated migration barriers for Al are consistent with the general trend of increasing migration barriers as ionic charge density increases, progressing from Li to Mg to Al. The lower barriers in ZZ--pris stacking suggest that Al ions can migrate more easily in this arrangement, supporting sustained capacity. Enhanced ion mobility is crucial for preserving capacity over prolonged cycling. VahidMohammadi~\textit{et al.}~\cite{vahidmohammadi2017two} reported a decline in capacity over repeated cycling from initial capacity of 335mAh~g$^{-1}$ to 112 mAh~g$^{-1}$ after 20 cycles, but did not investigate changes in stacking arrangement. Our results suggest that the structural transition from ZZ--pris to ZZ--oct stacking upon Al intercalation — which we find to be associated with substantially higher Al migration barriers - could contribute to this observed capacity fade. This offers a possible mechanistic explanation linking stacking changes to electrochemical performance degradation in Al-ion intercalated MXenes.

\subsection{Intercalated MXenes - Concentrated Limit}

\begin{figure}[ht]
    \centering
    \includegraphics[width=\columnwidth]{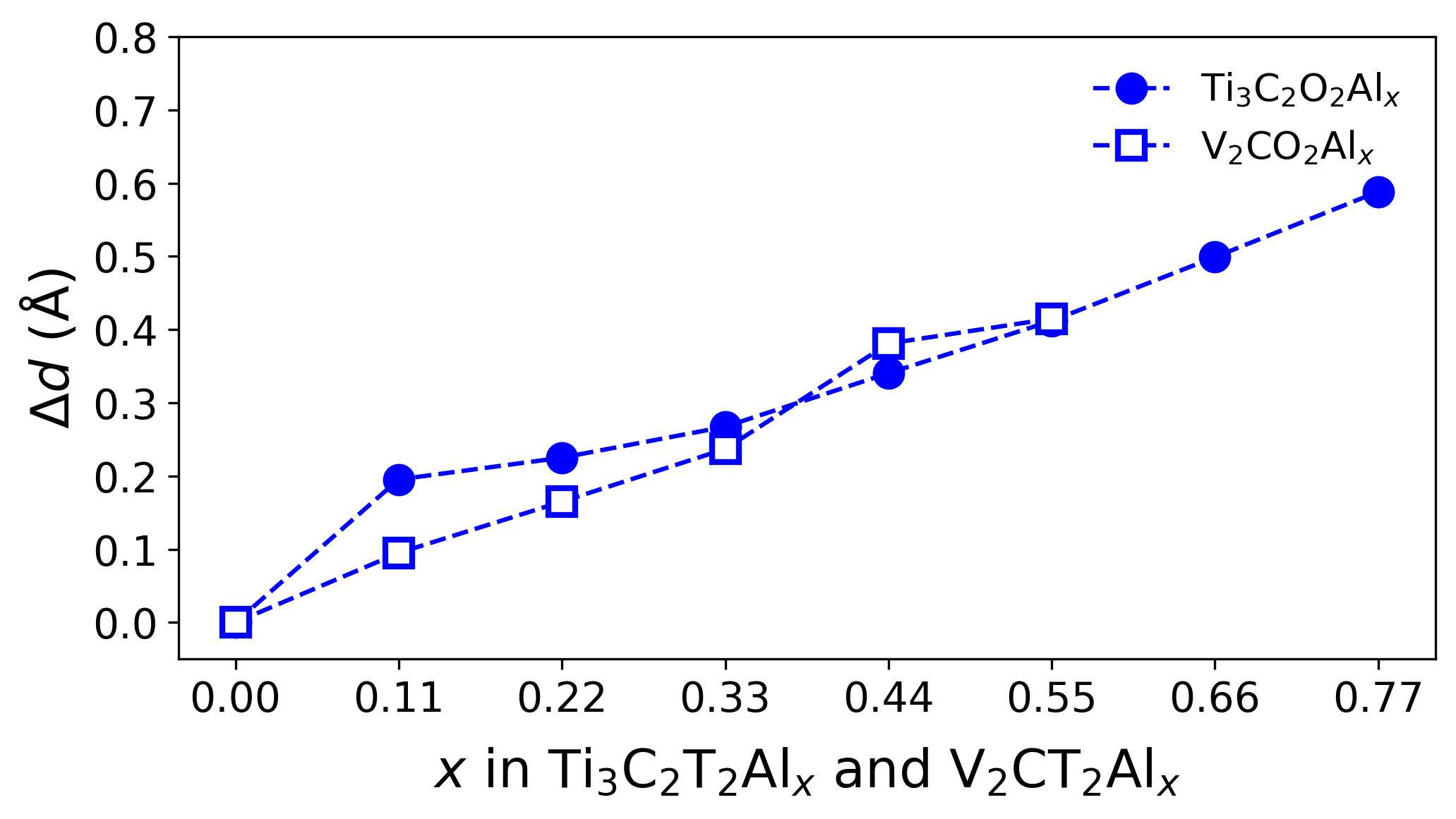}
    \caption{Change in interlayer spacing ($\Delta d$) with increasing Al concentration intercalated in ZZ--oct stacking. Solid circles represent \TiCO\ and open squares represent \VCO.}
    \label{Fig5}
\end{figure}

The change in the interlayer spacing, $\Delta d$, with increasing Al concentration in the ZZ–oct stacking is shown in Fig.~\ref{Fig5}. For both \TiCO\ and \VCO, $\Delta d$ increases approximately linearly with Al content, indicating a gradual expansion of the interlayer spacing as additional Al ions are accommodated. This behaviour is consistent with an enhanced electrostatic repulsion resulting from the higher concentration of intercalated Al.

\begin{figure}[ht]
    \centering
    \includegraphics[width=\columnwidth]{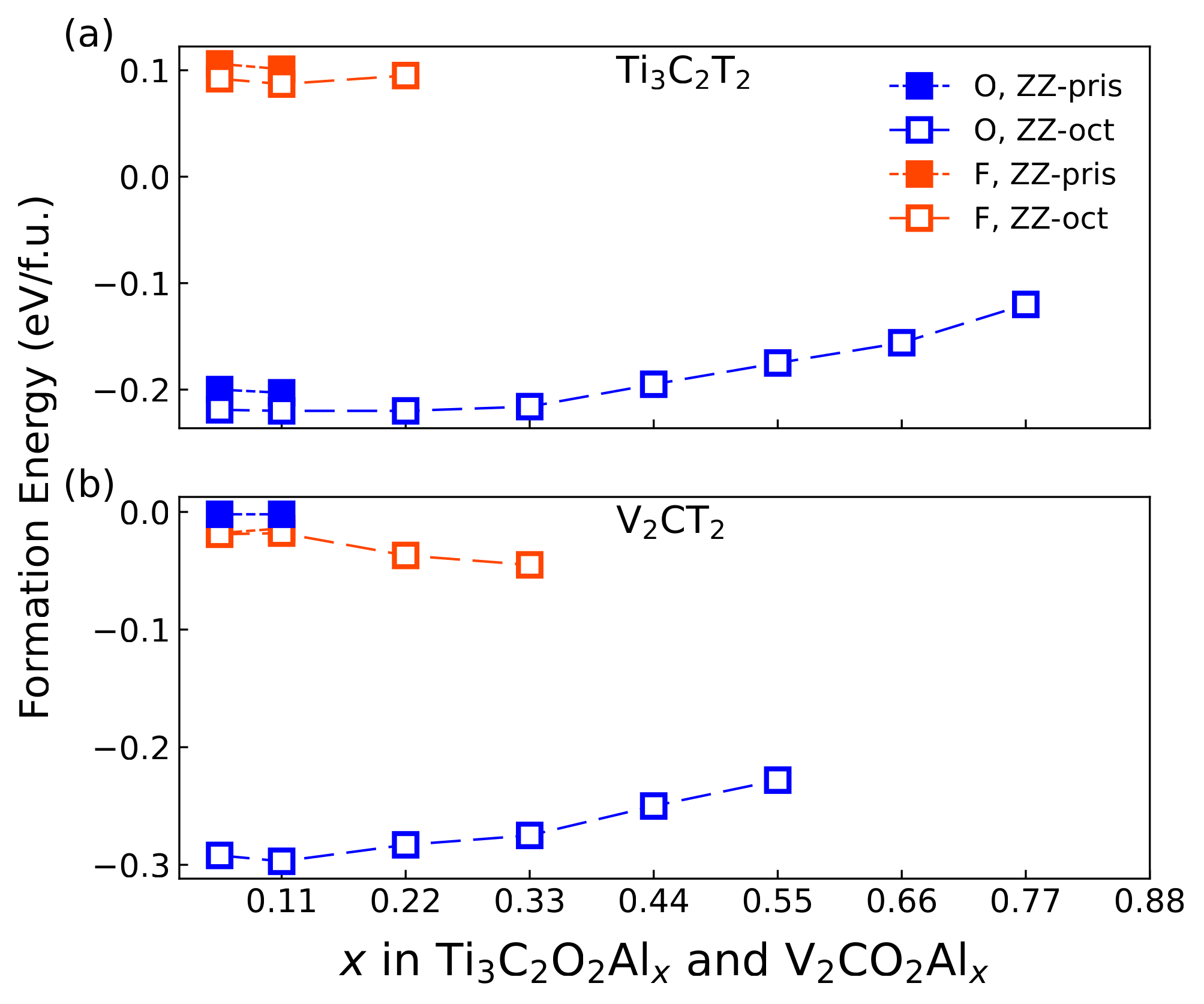}
    \caption{Formation energy (eV/f.u.)\ as a function of Al concentration for (a) Ti$_3$C$_2$T$_2$ and (b) V$_2$CT$_2$. ZZ--oct and ZZ--pris
stackings are denoted as open squares and solid squares, respectively. –O
terminations are shown as long dashed blue lines and –F
terminations as short dashed orange lines.}
    \label{Fig6}
\end{figure}

Fig.~\ref{Fig6} shows how the formation energy changes with increasing Al concentration. For \TiCO, Al intercalation remains favourable up to a maximum concentration of 0.77~Al/f.u., whereas for \VCO\ the stability limit is reduced to~0.55 Al/f.u. These predicted limits are in good agreement with experimental observations by VahidMohammadi~\textit{et al.}\cite{vahidmohammadi2017two}, who reported achievable Al concentrations of up to $\sim$0.5 Al/f.u. in multilayered V$_{2}$CT$_{x}$. Note that substantially higher intercalant concentrations have been reported previously for monovalent species, such as 2 Li/f.u. in Ti$_2$C~\cite{naguib2012mxene} and 2 Na/f.u. in Ti$_3$C$_2$~\cite{wang2015atomic}, highlighting the additional thermodynamic constraints associated with trivalent Al intercalation.

For F-terminated MXenes, the maximum Al concentration achievable before structural instability occurs are much lower, reaching only 0.33~Al/f.u.\ in \VCF\ and 0.22~Al/f.u.\ in \TiCF. Positive formation energies in the latter indicate that even these low concentrations are thermodynamically unfavorable. This reduced stability likely arises from the weaker and more ionic TM–F bond, which is more easily broken in favour of forming highly stable Al–F bonds. As Al concentration increases, this competition becomes more pronounced, leading to the removal of F from the MXene surface and the subsequent degradation of the functionalized layers. 

To further understand the chemical origins of these stability limits, we next investigate the redistribution of electronic charge upon Al intercalation.
Bader charge analysis~\cite{bader1981quantum} is used to understand the influence of intercalant type, MXene composition, and surface termination on charge distribution. Table~\ref{tab:Tab2} reports the total change in Bader charge for each atomic species upon intercalation, summed over all atoms of a given type. 
\begin{table}[h!]
\small
\centering
\caption{\ Total change in Bader charge per atomic species upon intercalation into a $3\times3$ MXene supercell, relative to the pristine MXene. TM(s) is the central transition metal atom in a MXene layer. TM(c) is the central transition metal atom in a MXene layer. T$_\mathrm{bond}$ and T$_\mathrm{unbond}$ indicates termination group atoms that are directly bonded and not bonded, respectively, to the intercalant. Positive values indicate electron gain, negative values indicate electron loss. Values are summed over all atoms of each type.}
\label{tab:Tab2}
\begin{tabular*}{0.48\textwidth}{@{\extracolsep{\fill}}lcccccc}
\hline
\hline
System & Intc. & TM(s) & C & TM(c) & 
T$_\mathrm{bond}$ & T$_\mathrm{unbond}$ \\
\hline
\multicolumn{7}{l}{\textbf{Ti$_3$C$_2$O$_2$}} \\
Al$_{0.055}$ & -2.40 & 0.29 & 0.16 & 0.11 & 1.70 & 0.18 \\
Al$_{0.77}$  & -2.44 & 10.94 & 4.49 & 1.93 & 16.81 & --- \\
Na$_{0.055}$ & -0.79 & -0.09 & -0.10 & -0.03 & 0.78 & 0.22 \\
Mg$_{0.055}$ & -1.66 & 0.09 & 0.14 & -0.06 & 1.30 & 0.21 \\
\hline
\multicolumn{7}{l}{\textbf{Ti$_3$C$_2$F$_2$}} \\
Al$_{0.055}$ & -2.49 & 1.23 & 0.06 & 0.13 & 0.80 & 0.27 \\
Al$_{0.22}$  & -2.50 & 5.40 & 1.28 & 0.40 & 2.47 & 0.46 \\
Na$_{0.055}$ & -0.80 & 0.34 & -0.09 & -0.04 & 0.40 & 0.18 \\
Mg$_{0.055}$ & -1.69 & 0.80 & -0.04 & 0.05 & 0.65 & 0.23 \\
\hline
\multicolumn{7}{l}{\textbf{V$_2$CO$_2$}} \\
Al$_{0.055}$ & -2.39 & 0.87 & -0.28 & --- & 1.43 & 0.37 \\
Al$_{0.55}$  & -2.46 & 8.54 & 1.80 & --- & 14.24 & --- \\
Na$_{0.055}$ & -0.79 & 0.48 & -0.10 & --- & 0.38 & 0.04 \\
Mg$_{0.055}$ & -1.67 & 0.67 & 0.03 & --- & 0.95 & 0.02 \\
\hline
\multicolumn{7}{l}{\textbf{V$_2$CF$_2$}} \\
Al$_{0.055}$ & -2.50 & 1.60 & 0.11 & --- & 0.68 & 0.11 \\
Al$_{0.33}$  & -2.14 & 8.95 & 1.60 & --- & 1.47 & 0.82 \\
Na$_{0.055}$ & -0.80 & 0.54 & 0.03 & --- & 0.19 & 0.05 \\
Mg$_{0.055}$ & -1.69 & 1.05 & 0.05 & --- & 0.48 & 0.10 \\
\hline
\hline
\end{tabular*}
\end{table}

As monovalent and divalent atoms, respectively, Na donates approximately 0.8~e per atom to the MXene layer, while Mg donates around 1.7~e per atom. Al donates approximately 2.4-2.5~e per atom, close to its nominal valence.

In the dilute Na intercalation limit, the charge donated to the MXene is small and largely localized, indicative of simple electrostatic adsorption polarizing the MXene charge density. For example, in \TiCO, the surface Ti atoms (TM(s)) and the C atoms also lose charge, while the majority of donated electrons are accepted by the bonded surface oxygen terminations (T$_\mathrm{bond}$). This weakly screened, surface-localized transfer is consistent with previous charge density difference and Bader analyses of Na intercalation into \TiC\ and \VC~\cite{caffrey2018effect}.

For Mg intercalation, the MXene framework (TM(s) and C) begins to gain charge, indicating that the slab now acts as a modest electron acceptor, though the surface terminations remain the primary charge sinks.

For Al at the same dilute concentration (Al$_{0.055}$), the TM(s) atoms gain significantly more charge than for Na or Mg. In this limit, the donated electrons are predominantly accepted by the TM(s) and termination atoms, with T$_\mathrm{bond}$ receiving substantially more than T$_\mathrm{unbond}$. The C and TM(c) atoms hold just 12\% of the transferred charge.

At higher Al loadings, the total charge gained by TM(s) atoms grows nonlinearly. For instance, TM(s) atoms gain 10.94~e in \TiCO Al$_{0.77}$ compared to 0.29~e in Al$_{0.055}$, nearly three times larger than a simple sum might suggest. At these concentrations, a larger fraction of the charge is donated to TM(s) relative to the oxygen sites. At low concentrations $\sim$80\% of the charge transferred to \TiCO\ resides on the terminating oxygen atoms and just over 10\% on TM(s). At high concentrations, up to $\sim$30\% of the charge is donated to TM(s), 50\% to T$_\mathrm{bond}$, while the C and TM(c) atoms take $\sim$20\%.

F-terminated MXenes, by contrast, concentrate a larger fraction of the donated electrons on the surface TM(s), with the F atoms accepting a smaller fraction.
For example, in low Al loadings, 64\% of the transferred charge resides on V atoms in \VCF\ compared to 36\% in \VCO. In \TiCF, 50\% of the transferred charge is on the V atoms, whereas is it is only 12\% in \VCO. Higher loadings reduce the fraction on terminating F atoms in \TiCF\ from 43\% to 30\%, while the fraction on carbon increases from 2\% to 13\%.

The stronger localization of charge on surface transition metals in F-terminated MXenes weakens TM-F bonds and is consistent with their reduced thermodynamic stability upon Al intercalation, while the more distributed charge acceptance in O-terminated systems supports higher stable Al concentrations.

\begin{figure}
    \centering
    \includegraphics[trim=-10 -10 -10 -10, width=\columnwidth]{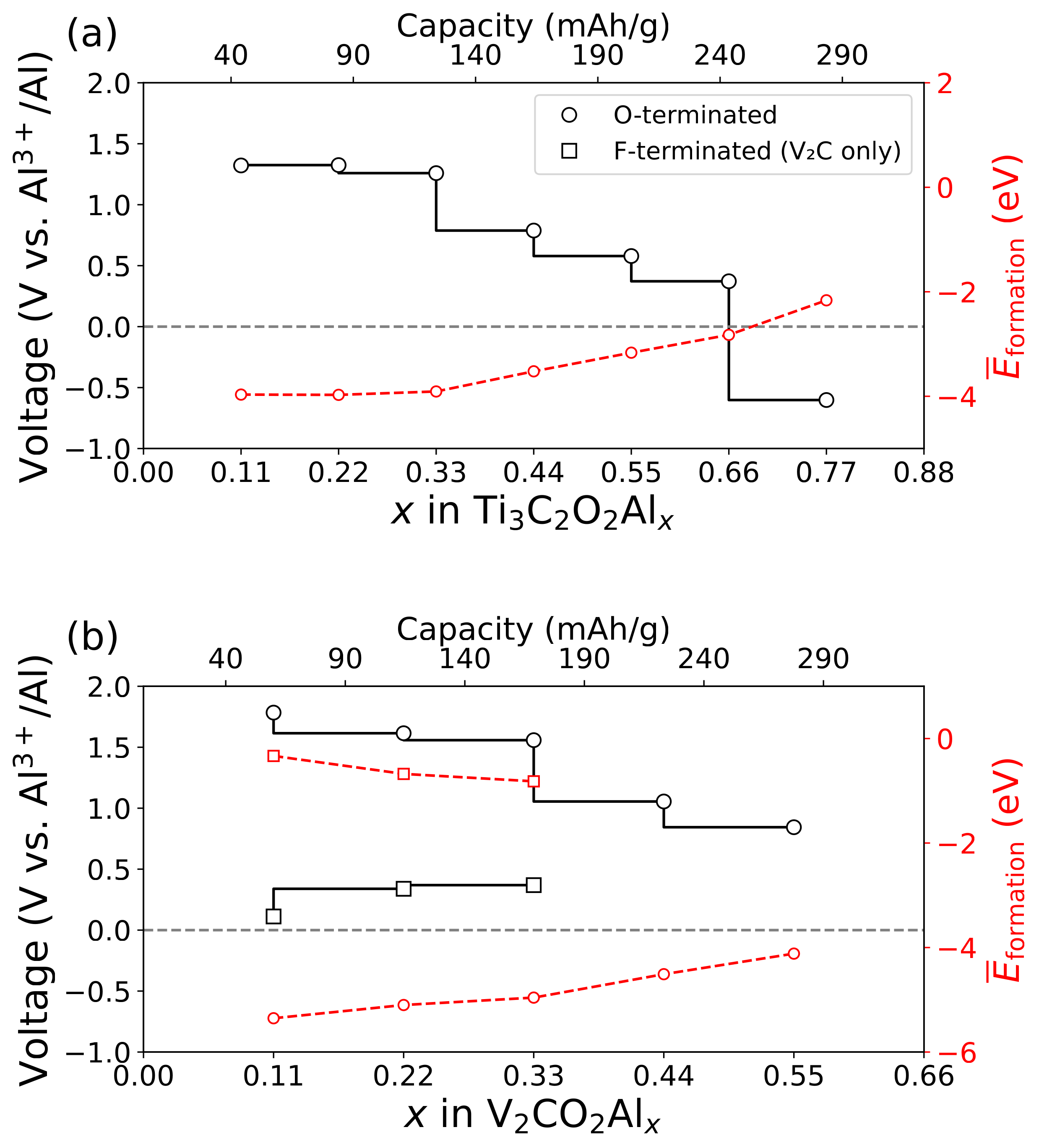}
    \caption{Open circuit voltage (V), capacity (mAh/g) and average formation energy (eV) for ZZ-oct stacked MXenes with increasing Al concentration intercalated in (a) \TiC\ and (b) \VC. O-terminated structures are indicated with circles. F-terminated structures are represented by squares.}
    \label{Fig7}
\end{figure}

Finally, the open circuit voltages (OCVs) and capacities of the two MXenes with --O and --F terminations are shown in Fig.~\ref{Fig7}. 
For \TiCO, the OCV ranges from 1.3~V to -0.6~V with increasing Al concentration, with positive OCV values indicating spontaneous intercalation of Al. The OCV becomes negative for \TiCO Al$_{0.77}$ indicating structural instability at that point, corresponding to a maximum theoretical capacity of 283.48~mAh/g.
The OCV of \VCO\ is higher that than of \TiCO\ by almost 0.5~eV at the lowest concentration of Al considered. The difference reduces to a constant 0.3~eV at higher concentrations. 

The OCV for \VCO\ and \VCF\ is also positive over the entire range of Al concentrations considered, although with considerably lower values for the F-terminated structure. For \VCO, it ranges from 1.8~V to 0.8~V with increasing Al concentration up to \VCO Al$_{0.55}$, while for \VCF, it increases from 0.1~V at \VCF Al$_{0.11}$ to 0.4~V at \VCF Al$_{0.33}$. For \VCO, the maximum capacity is 277.63~mAh/g, while for \VCF, is is reduced to 166.51~mAh/g. These lower values are consistent with the reduced thermodynamic stability and the surface-localized charge transfer observed for F--terminations.

Previous experimental measurements report an OCV range of 0.1–1.8~V and a capacity of approximately 335~mAh/g for \VC~\cite{vahidmohammadi2017two}, slightly higher than our calculated values. The differences can be attributed to the idealized conditions in the calculations, including the absence of co-intercalated species and finite-temperature effects.

\section*{Conclusions}
In this work, we examined the role of stacking configurations, surface terminations, and intercalant coordination in \TiC\ and \VC\ MXenes to determine their potential as cathode materials in Al-ion batteries.

We found that, upon Al intercalation, an octahedral stacking of the MXene layers becomes strongly energetically favourable, in contrast to Li intercalation which has previously been shown to stabilise a prismatic stacking. 
We find that Al intercalation into O-terminated structures is always thermodynamically favourable, particularly for the ZZ-oct stacking. In contrast, Al intercalation into \TiCF\ is unfavourable and only marginally favourable in \VCF. 

We also found that both the nature of the surface terminations and the stacking configurations play a crucial role in determining the interlayer distance of MXenes after intercalation, which has a direct impact on possible structural degradation. In general, ZZ--oct stacking resulted in smaller interlayer expansions compared to ZZ--pris stacking for all intercalants considered. Likewise, O-terminated MXenes consistently exhibited smaller interlayer expansions than their F-terminated counterparts across all intercalants. In particular, Al intercalation in \VCO\ resulted in an exceptionally low interlayer expansion of approximately 0.1~\AA, in excellent agreement with previous experimental observations, and in support of that work's supposition that bare Al was intercalated rather than chloroaluminates like~$[\mathrm{AlCl}_4]^-$~and~$[\mathrm{Al}_2\mathrm{Cl}_7]^-$
. 
In contrast, the combination of  F-terminations and ZZ--pris stacking produced the largest interlayer distance change. 

However, while we find that the ZZ--oct stacking is thermodynamically the most stable, this structure hampers ion mobility. Migration barrier analysis revealed that ZZ--pris stacking results in lower migration barriers, indicating a trade-off between structural stability and ion transport. 

We also show that O-terminated MXenes tend to localize charge on the termination groups, reducing transfer to the metal atoms compared to the F-terminated structures which promote charge transfer to the transition atoms.

We find that \TiCO\ can achieve a high theoretical capacity of 283.48~mAh/g at 0.77~Al/f.u. with an OCV range of 1.3 to -0.6~V. For \VCO, a maximum intercalation concentration of 0.55~Al/f.u. is achievable, limiting the capacity to 277.63~mAh/g with an OCV in the range of 1.7 to 0.8~V, in relatively good agreement with previous experimental work.

In summary, both the stacking type and surface termination significantly influence MXene performance. Although ZZ--oct stacking favours intercalant stability, stabilising the ZZ--pris stacking would enable enhanced ion mobility. 
Overall, the interlayer distance in MXenes is influenced by both the stacking type and the termination type. To utilise MXenes effectively as battery electrodes, stable stacking configurations that promote high ion mobility are essential for enhanced ion transport and increased capacity.

\section*{Conflicts of interest}
There are no conflicts to declare.

\section*{Data availability}
Data for this article, including structural files, and input and output files for density functional theory calculations, are available in the NOMAD repository~\cite{Scheidgen2023} at https://doi.org/10.17172/NOMAD/2026.02.12-1.

\section*{Acknowledgements}

This work was supported by an UCD Ad Astra fellowship.
Computational resources were provided by the supercomputer
facilities from Irish Centre for High-End Computing (ICHEC) in partnership with LuxProvide.

\balance

\bibliography{rsc} 
\bibliographystyle{rsc} 
\end{document}